\documentclass[aps,prl,preprint,groupedaddress]{revtex4-1}

\usepackage{graphicx}
\usepackage{epstopdf}
\usepackage{bm}
\begin{document}


\title{Combination of $e^+$/$e^-$ ratio from  AMS-02 and gamma ray line from Fermi-LAT with implication for
Dark Matter}


\author{
Shiyuan Li$^{1}$
}
\author{
Yonggang Luo$^{1}$
}
\email{
luoyg@mail.sdu.edu.cn
}
\affiliation{
$^1$ Taishan College, Shandong University, Jinan, 250100, China \\
}


\date{\today}

\begin{abstract}
The precise AMS-02 data provide definite information for the $e^+$/$e^-$ ratio in 100 - 350 GeV region. Assuming that the recent gamma ray line observed by Fermi-LAT experiment is product of dark matter in space and taken as input. We make a global fit for the AMS-02 $e^+$/$e^-$ spectrum with both dark matter and pulsar contribution. For this spectrum over 130 GeV, pulsar is the dominant contribution. We also investigate the constrains on hadron production from dark matter annihilation.
\end{abstract}


\maketitle


\section{I  INTRODUCTION}

Dark Matter (DM) is  introduced based on traditional astronomy observation as well as the cosmology global fitting \cite{bib:Zwicky}.
For this aspect, only the gravitational  interaction is relevant. For its microscopic structure and its particle properties, including  the interactions beyond gravity it proceed,
other experiments are designed such as underground direct detection and the cosmic ray experiments to detect its decay/annihilation products.
There have been many groups devoted to this purpose, of them the most up to data are
AMS-02\cite{bib:Aguilar} and Fermi-LAT\cite{bib:Fermi,bib:Ackermann} experiments.

%

The experiment data, including PAMELA\cite{bib:OAdriani}, Fermi-LAT\cite{bib:Ackermann} and AMS-02\cite{bib:Aguilar}, had indicated a significant excess above the standard Galactic cosmic ray expectations.
In order to explain it, several possible scenarios, such as DM\cite{bib:Lei}, pulsars\cite{bib:profumo,bib:FL},
 have been proposed. It is also possible to fit this data by both DM and pulsars \cite{bib:Feng}. An interesting result from Fermi-LAT gamma ray line around 135 GeV from the analysis\cite{bib:Fermi,bib:Weniger,bib:Rajaraman,bib:ARajaraman,bib:TBringmann,bib:Hektor,bib:Su,bib:Whiteson,bib:DPFinkbeiner,bib:AHektor}
can be used as input for the DM particle mass, if one take that the gamma ray and $e^+e^-$ pair are products from annihilation of the same particle \cite{bib:Feng}. The difficulty for
this way of combination fitting is that the relative contribution from DM and pulsars are almost free parameters since the PAMELA data only extent to 100 GeV.




The work\cite{bib:Feng} only fit the PAMELA data with this input. If one takes the DM particle is around  $n$ $\times$ 135 GeV, where $n$ is the number of the produced electron per DM particle,
its contribution decreases to vanish fast beyond this energy.
The PAMELA data can not provide information beyond 100 GeV to give restriction to pulsars.
However, the AMS-02 data provide the information up to 350 GeV. Furthermore, a  fine structure at around 135 GeV seems existed and are helpful to assume that the DM particle mass is 135 GeV.
So we do the fitting as the following way: First to fit the data above 135 GeV by only using pulsars,
so that make good constrain on the pulsars contribution;
Then adding the electron flux produced from DM particle annihilation process, to look for the corresponding DM particle parameters.

The results show that we can employ this two component way to explain the data well. This again confirms that the relation of the $e^{+}$/$e^{-}$ excess with the gamma ray data can be assumed. In this step
the question is what kind of mechanism to relate the electron and gamma ray production.

1) If the gamma ray is really a line. Two DM particles annihilate into two photons or two electrons at exactly 135 GeV.

2) If the gamma ray data show a narrow resonance. One can assume that the photons and electrons are from a mediate resonance which is massive.
On the one hand, such kind of models are widely discussed\cite{bib:Cholis,bib:Douglas}. On the other hand, this massive mediate particle can lead
   directly and indirectly quarks hence hadron productions\cite{bib:Evoli}. So what kind of constrain can be deduced from the up to date antiproton data need to be investigated.

Our analysis shows that the up to date precision will not lead strict constrain on the cross section of the DM particles annihilating to quark sector. So a large group of models\cite{bib:Cholis,bib:Douglas} can be investigated based on
these relations and more observation about a new massive particle can be predicted.


The outline of this paper is as follows: In section II, we discuss the contribution of three components including background, pulsar and dark matter to AMS-02 spectrum.
In section III, we investigate the antiproton/proton ratio with input from the fitting result in $e+$/$e^-$ ratio. In section IV, we discuss and conclude our results.

\section{II Combination on AMS-02 spectrum with DM and pulsars contributions}
In this section, we investigate the contribution from pulsars and DM particle to the $e^+$/$e^-$ excess employing the precise AMS-02 data, especially its fine structure of the spectrum around 100 GeV. Assuming the gamma ray excess around 135 GeV\cite{bib:albert}, and the positrons are produced from the same kind of DM particle annihilation/decay.

In order to calculate the background electron flux, we employ the Galprop code\cite{bib:Andrew}, which is a numerical code for
calculating the propagation of relativistic charged particles and the diffuse emissions produced during their propagation.
In this paper, all the parameters value were taken from the best-fit Bayesian model of Trotta et al. (2011)\cite{bib:Trotta} which is showed in Table I.

\begin{table}
\centering
\begin{ruledtabular}
\begin{tabular}{lcccccc}
\hline
$D_{0}$ &$\delta$ &$v_{alf}$ &$z_{h}$ &$v_{1}$ &$v_{2}$ &$N_{p}$\\ \hline
6.59 &0.30 &39.2 &3.9 &1.91 &2.40 &5.00\\
\end{tabular}
\caption{Summary of Input Parameters}
\end{ruledtabular}
\end{table}

Considering the fact that the gamma ray excess emerged around the 135 GeV, we
can know that there is no effect on positron flux above 135 GeV region by DM annihilation.
This means that the excess positrons in this region are produced only from pulsars. The relevant electron flux from pulsars are given by
\begin{center}
\begin{equation}
Flux = AE^{B}e^{-E/C}\rho
\end{equation}
\end{center}
where $E$ is energy and $\rho$ is given by
\begin{center}
\begin{equation}
\rho = (\frac{r}{8.5})^{2.35}e^{-5.56 (\frac{r-8.5}{8.5})}e^{-5.56\left| \frac{z}{0.2} \right|}
\end{equation}
\end{center}
In this formula above, there are three parameters $A$, $B$ and $C$ should be determined by fitting AMS-02 data.
In this work, we employ data above 135 GeV to fit. We find that, for our background model,
the best fitting parameters value are $A$ = $2.1\times10^{-27}$, $B$ = -1.28 and $C$ = $4\times10^{5}$. In \cite{bib:profumo}, the parameters values are determined as B = -1.9 and C = $2\times10^{6}$ which adopts a rescaling of the Galprop diffuse electron flux by a factor 0.8 in order to fit the global $e^+$/$e^-$ spectrum from 5 GeV to 350 GeV.
The $e^+$/$e^-$ spectrum, which combine the background and products from pulsars (Fig. 1.), shows that there are free space for DM below the 135 GeV.

\begin{figure}[t]
  \centering
  \includegraphics[width=0.8\textwidth]{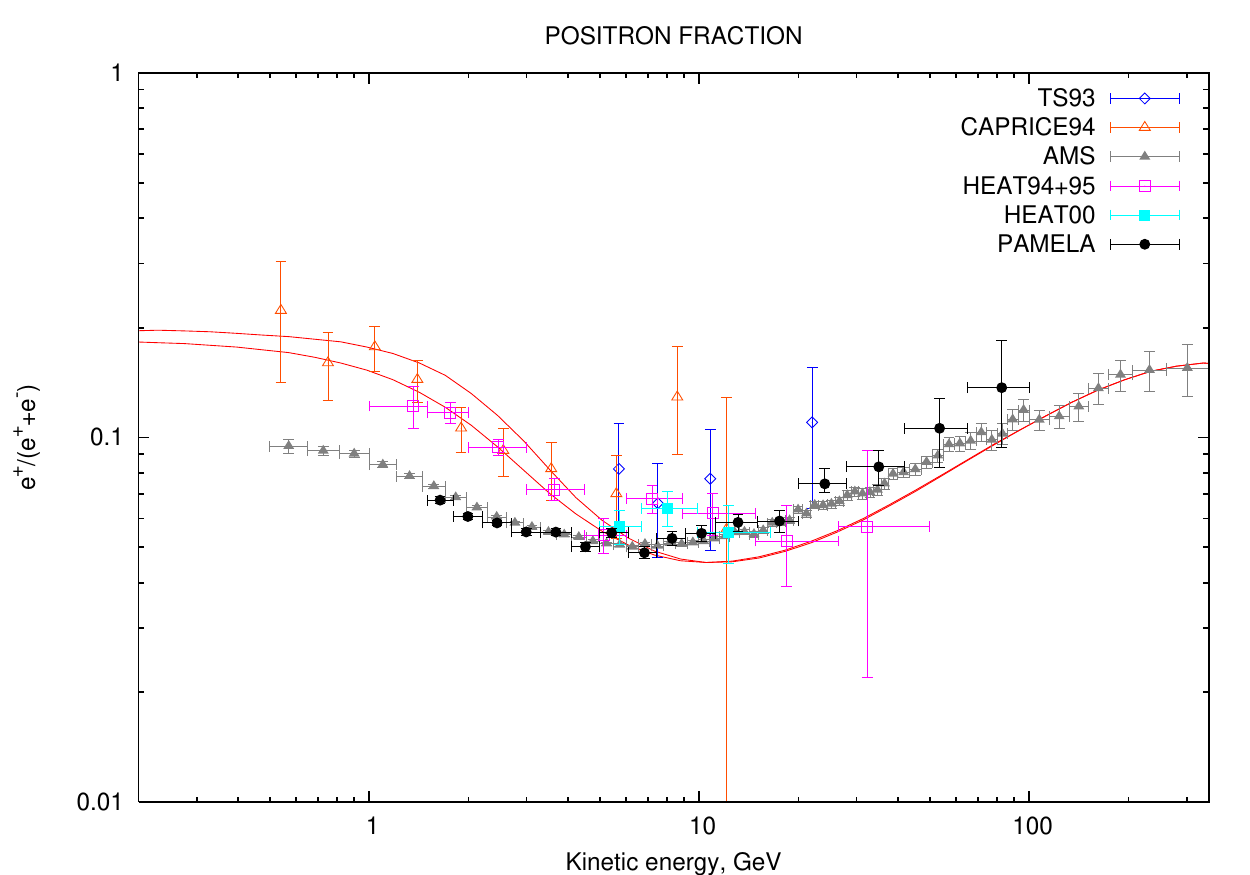}
  \caption{$e^+$/$e^-$ ratio by only adding the pulsars contributions to the background}
  \label{2}
 \end{figure}

The products from DM annihilation can be added into the $e^+$/$e^-$ spectrum,
The fine structure of the AMS-02 spectrum around 100 GeV results from adding DM contribution.
The relevant electron flux from DM annihilation are given by
\begin{center}
\begin{equation}
Flux = \frac{\rho^{2}}{m^{2}}\frac{dN}{dE}\left\langle \sigma v \right\rangle
\end{equation}
\end{center}
where $m$ is the DM mass, $\frac{dN}{dE}$ is normalized for one time of annihilation, $\left\langle \sigma v \right\rangle$ is the pair annihilation rate and $\rho$ is the density of DM particles. In this work, we consider three kinds of DM distribution profiles, which are named as isothermal profile, Evans profile and alternative profile in Galprop code\cite{bib:Andrew}.

In equation (3), we employ a Gaussian distribution with $E_c = 135$ GeV for $\frac{dN}{dE}$. If we take Fermi-LAT data as input, $m = 2E_c$ for process $\chi\chi$ $\rightarrow$ $e^{+}$$e^{+}$$e^{-}$$e^{-}$ and $m = E_c$ for process $\chi\chi$ $\rightarrow$ $e^{+}$$e^{-}$. $\left\langle \sigma v \right\rangle$ is fitted to AMS-02 data from 8 GeV to 132 GeV. Fig. 2, Fig. 3 and Fig. 4. correspond the three DM density profile: isothermal profile, Evans profile and alternative profile.
The best fit values of $\left\langle \sigma\nu \right\rangle$ are showed in Table II.

\begin{table}
\centering
\begin{ruledtabular}
\begin{tabular}{lccc}
\hline
 &isothermal profile &evans profile &alternative profile\\ \hline
$\chi\chi$ $\rightarrow$ $e^{+}$$e^{+}$$e^{-}$$e^{-}$ &3.4 $\times$ $10^{-25}$ &3.5 $\times$ $10^{-24}$ &3.5 $\times$ $10^{-25}$ \\
$\chi\chi$ $\rightarrow$ $e^{+}$$e^{-}$ &1.7 $\times$ $10^{-25}$ &1.8 $\times$ $10^{-24}$ &1.7 $\times$ $10^{-25}$ \\
\end{tabular}
\caption{The best values of $\left\langle \sigma v \right\rangle$ ($cm^{3}/s$)for different density profiles}
\end{ruledtabular}
\end{table}

\begin{figure}[t]
  \centering
  \includegraphics[width=0.8\textwidth]{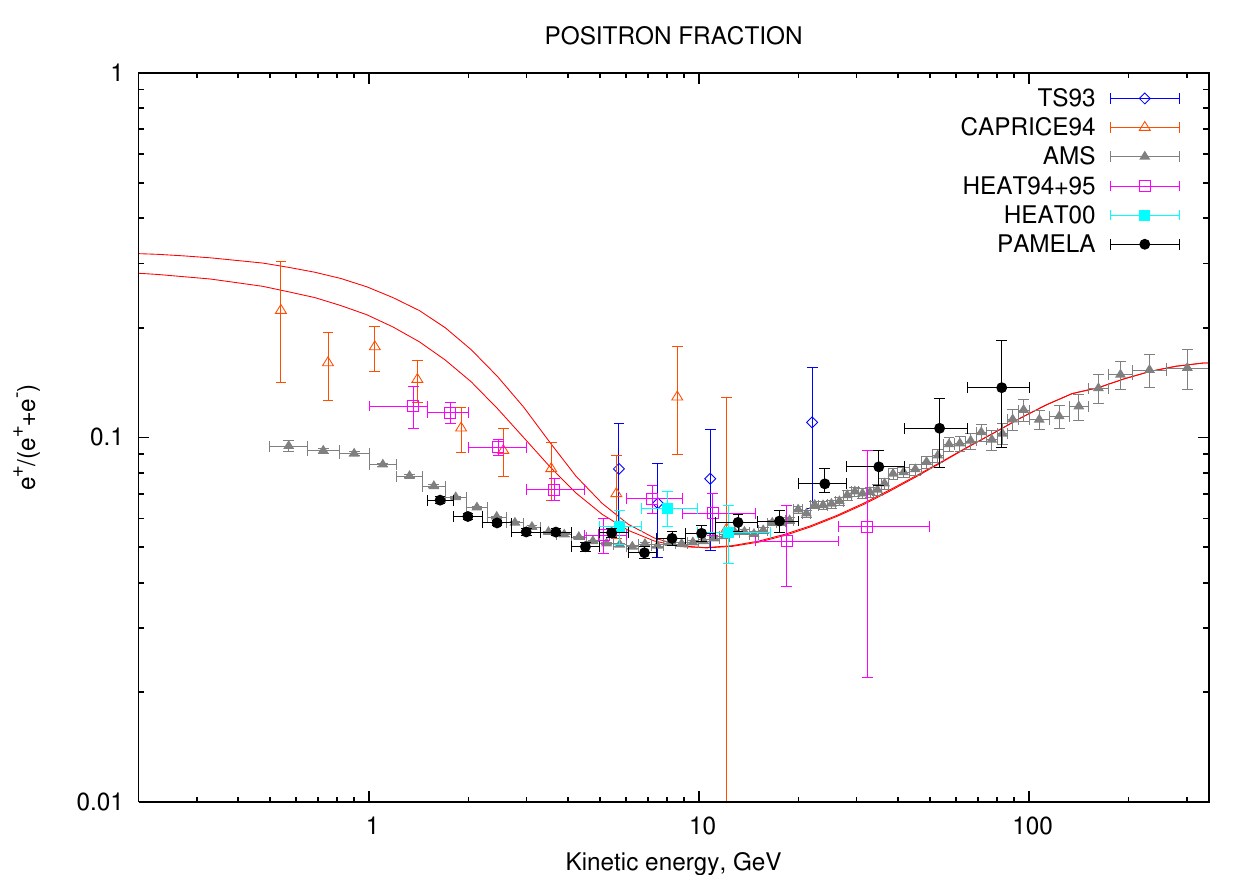}
  \caption{$e^+$/$e^-$ ratio for isothermal profile with $\left\langle \sigma v \right\rangle$ = 3.4 $\times$ $10^{-25}$ $cm^{3}/s$}
  \label{3}
 \end{figure}

\begin{figure}[t]
  \centering
  \includegraphics[width=0.8\textwidth]{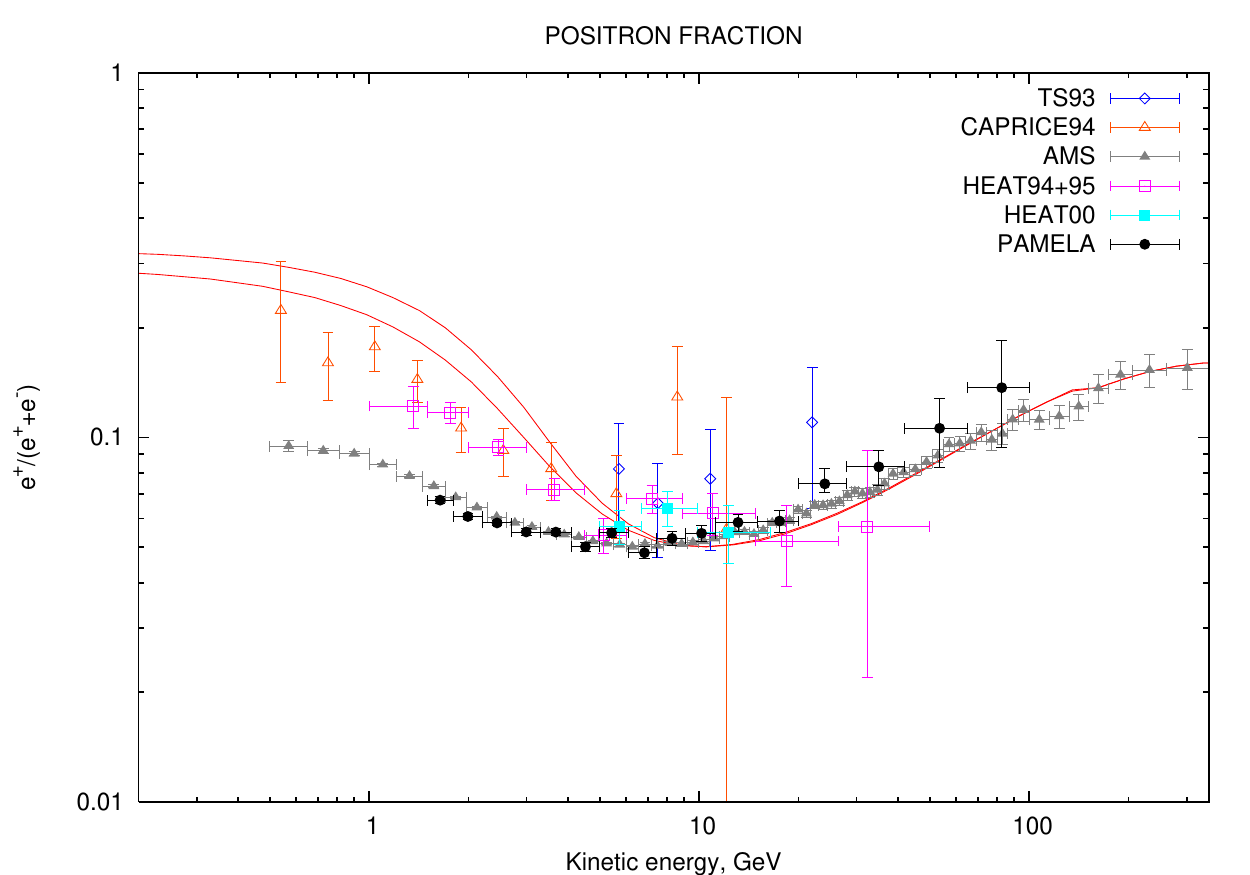}
  \caption{$e^+$/$e^-$ ratio for Evans profile with $\left\langle \sigma v \right\rangle$ = 3.4 $\times$ $10^{-25}$ $cm^{3}/s$}
  \label{4}
 \end{figure}

\begin{figure}[t]
  \centering
  \includegraphics[width=0.8\textwidth]{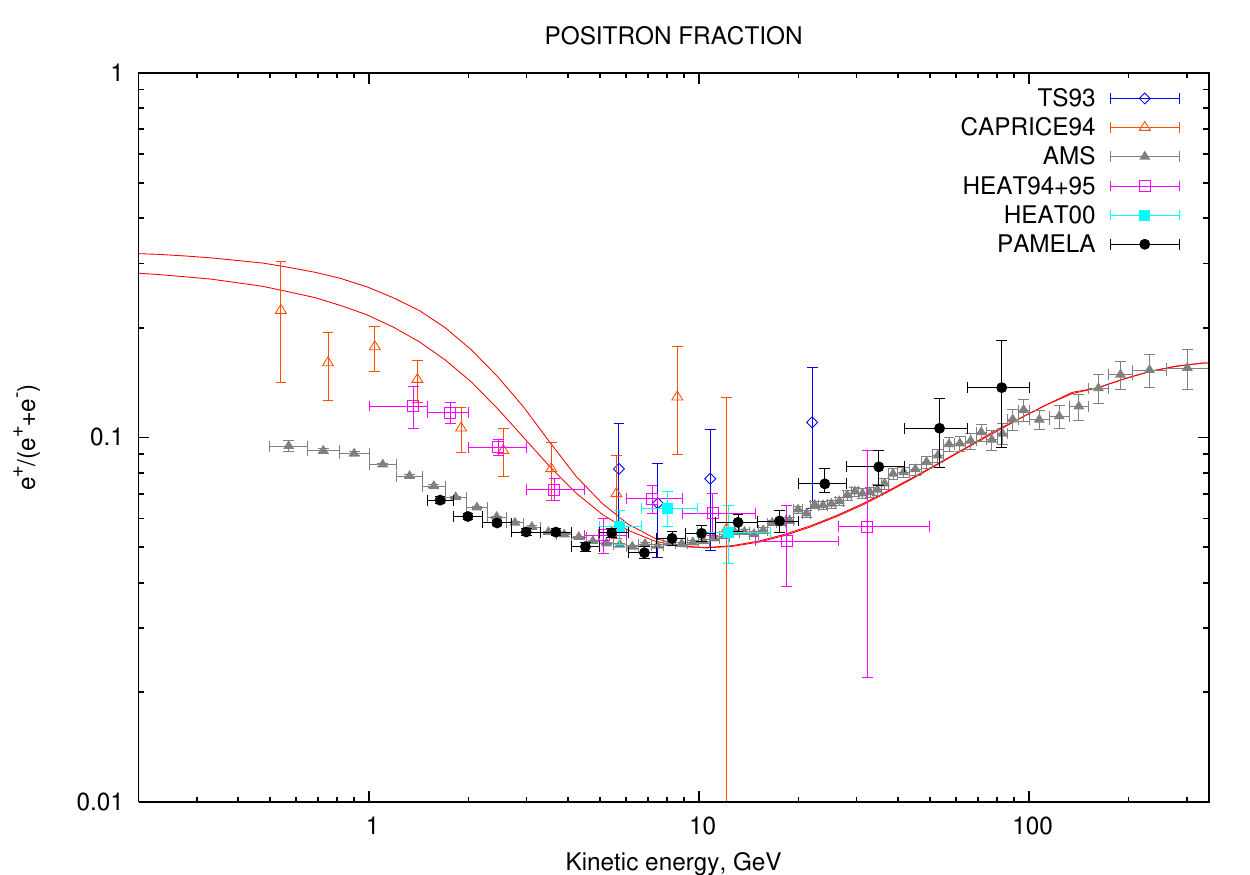}
  \caption{$e^+$/$e^-$ ratio for alternative profile with $\left\langle \sigma v \right\rangle$ = 3.5 $\times$ $10^{-24}$ $cm^{3}/s$}
  \label{5}
 \end{figure}

\section{III $\bar{p}$/$p$ ratio}

To further explore the property of the DM particles which can produce both electron and gamma ray, it is crucial to investigate more on the constrain of the cross section of the DM particles annihilate
to other kind of elementary particles. The above process are in principle direct or indirect electroweak interactions, or else the gamma and electron can not produce.
The other particles which can also take part in the electroweak interactions are quarks. So DM particles will, at least, indirectly produce quarks hence hadrons.
There are many DM models(e.g. \cite{bib:Cholis,bib:Douglas}), which predict a mediate scalar which can couple electrons and other leptons via the electroweak interactions. If it also couple to quarks,
the antiproton data will give a strong constrain to its property. When the scalar invariant mass smaller than two times proton mass, the antiproton and proton production rate can be suppressed. However the production of them can not be completely eliminated
since the scalar is mediate particle while DM particle mass is much larger than proton mass.

In the following, we assume that the production of quarks has the cross section of same order as that of lepton (in numerical calculations, we take exactly the same value), to see whether the antiproton data will
eliminate such a possibility.
 PAMELA and other experiments have  observed antiproton flux.
These data are considered consistent with background. We calculate the antiproton production rate and compare it with the error of data to see whether the above mechanism will give an
observable correction to those data.

One  key point  is that the probability of  the quark fragmenting  to a proton/antiproton is rather small, especially those with energy comparable with the original quark.
The fragmentation function at the scale of around $100$ GeV is shown in Fig. 5. (normalized for one event, $dN/dE$, vertical should be changed ).
This is calculated with help of Pythia. The Pythia code is employed for the following numerical calculation for $\frac{dN}{dE}$ input.

We employed  Galprop code above to calculate the evolution of antiproton and proton. In the numenrical calculation,
we take ${\left\langle \sigma v \right\rangle}_{\chi\chi \rightarrow q\bar{q}}$ equals to ${\left\langle \sigma v \right\rangle}_{\chi\chi \rightarrow l\bar{l}}$.
We use XDM model as an example $\chi\chi$ $\rightarrow$ $\phi$$\phi$ $\rightarrow$ $e^{+}$$e^{+}$$e^{-}$$e^{-}$\cite{bib:Cholis,bib:Douglas},
 so that it is a two particles to four particles process.

Using the value of $\frac{dN}{dE}$, $\left\langle \sigma v \right\rangle$ and $m$ for different DM density profile,
the antiprotons produced from DM annihilation process can be found by employing Galprop code. FIG. 5 shows that comparison between antiproton/proton ratio data and the theory value for three profiles and Fig. 7  shows that comparison between antiproton/proton ratio data error and the theory value for three profiles. Obviously, for isothermal profile and alternative profile, the quantity of antiprotons produced from Dark Matter below the experiment error, as Fig. 6 and Fig. 7, which means, currently, the cosmic ray experiments cannot find the excess antiproton even though DM actually produce antiproton as we expect. However, for Evans profile,
antiprotons from 6 GeV to 30 GeV are comparable to the error of data, as Fig. 6. and Fig. 7.

Therefore, the mediate state $\phi$ can couple with quarks and the mass of $\phi$ is not a crucial factor.
From the data analysis, we can estimated that the scalar particle is massive and comparable with the DM, if assuming the gamma ray is from a narrow resonance. We use the half
width to estimate the mass.

\begin{figure}[t]
  \centering
  \includegraphics[width=0.8\textwidth]{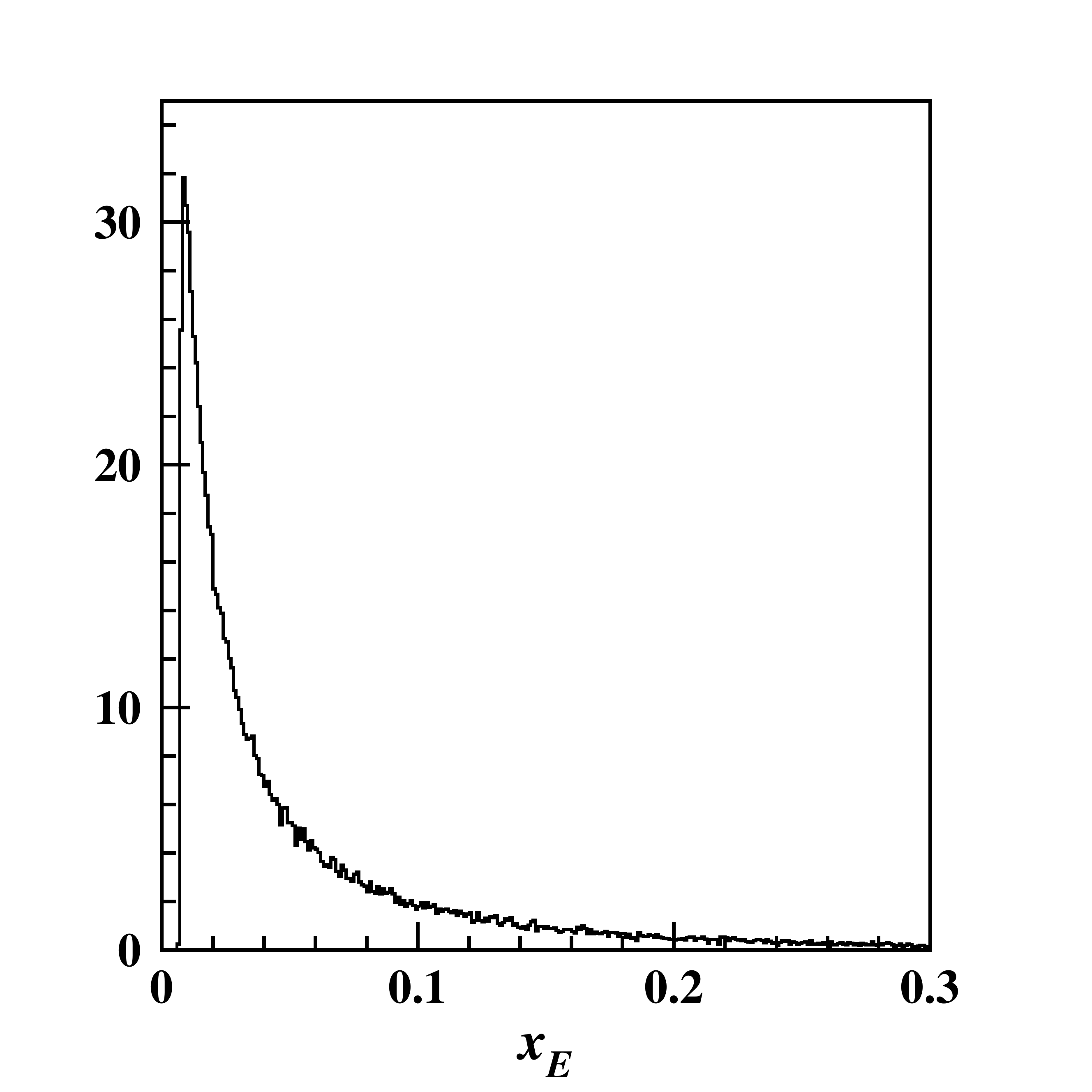}
  \caption{$\frac{dN}{dE}$ of antiproton from the quarks fragmentation. It is normalized to one quark.}
  \label{6}
 \end{figure}

\begin{figure}[t]
  \centering
  \includegraphics[width=0.8\textwidth]{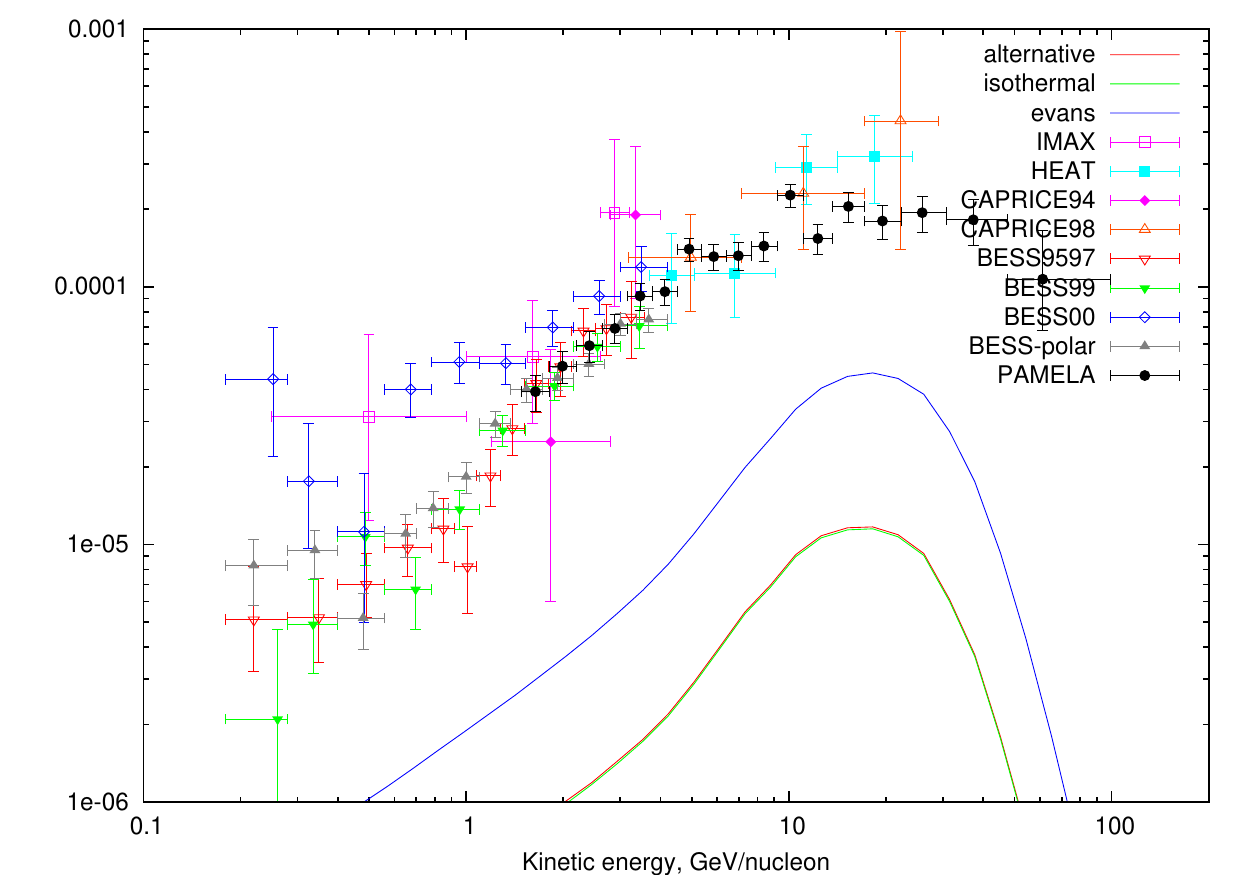}
  \caption{The dots express the antiproton/proton experiment data, and the lines express the theory value in our model for different profile}
  \label{7}
 \end{figure}

\begin{figure}[t]
  \centering
  \includegraphics[width=0.8\textwidth]{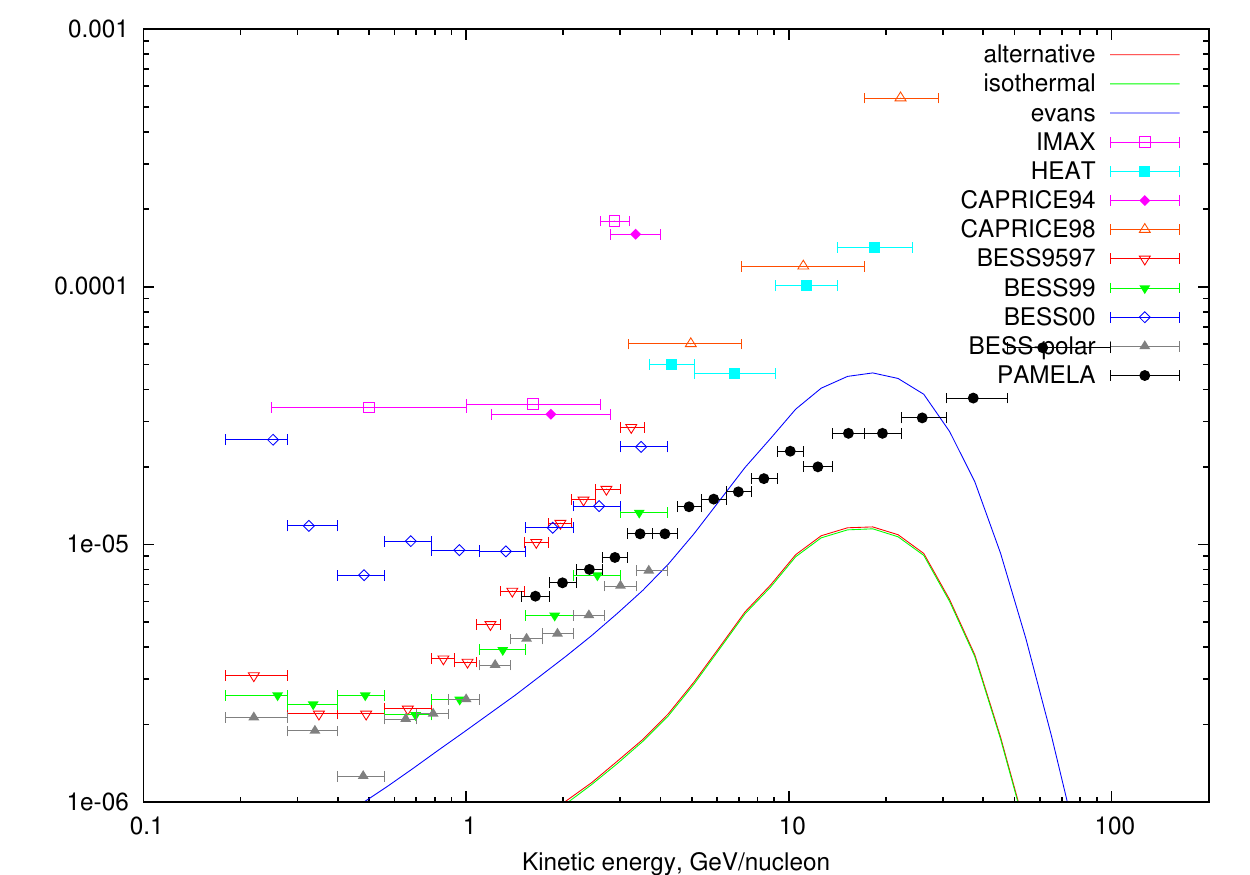}
  \caption{The dots express the antiproton/proton experiment error, and the lines express the theory value in our model for different profile}
  \label{8}
 \end{figure}

\section{IV Summary}
In this paper, we make a fitting for AMS-02 spectrum by using three components including background, pulsars and DM. In addition, we calculate the antiproton/proton ratio with DM contributions and compare it with the current experimental antiproton flux bound. For some specific density profile, the quantity of antiprotons produced from Dark Matter below the experiment error.

\end{document}